\documentclass[
 reprint,
superscriptaddress,
 amsmath,amssymb,
 aps,bbm
pra,
]{revtex4-1}
\usepackage{comment}
\usepackage{graphicx}
\usepackage{dcolumn} 
\usepackage{bm}
\usepackage{graphics}
\usepackage{bbold}
\usepackage{bbm}
\usepackage{amsmath,amsthm}
\usepackage{lipsum}
\usepackage{amssymb}
\usepackage{color}

\begin{document}
\title{Parallel-in-time optical simulation of history states}

\author{D.\ Pab\'on}
\affiliation{Departamento de F\'isica, FCEyN, Universidad de Buenos Aires, Pabell\'on 1, Ciudad Universitaria,  Buenos Aires (1428), Argentina}
\affiliation{Consejo Nacional de Investigaciones Cient\'ificas y T\'ecnicas (CONICET), Argentina.}

\author{L.\ Reb\'on}
\email[]{rebon@fisica.unlp.edu.ar}
\affiliation{Departamento de F\'{\i}sica, FCE, Universidad Nacional de La Plata, C.C. 67, La Plata (1900), Argentina}
\affiliation{Instituto de F\'isica de La Plata, UNLP - CONICET, Argentina}

\author{S. Bordakevich}
\affiliation{Departamento de F\'isica, FCEyN, Universidad de Buenos Aires, Pabell\'on 1, Ciudad Universitaria,  Buenos Aires (1428), Argentina}

\author{N.\ Gigena}
\affiliation{Departamento de F\'{\i}sica, FCE, Universidad Nacional de La Plata, C.C. 67, La Plata (1900), Argentina}
\affiliation{Instituto de F\'isica de La Plata, UNLP - CONICET, Argentina}

\author{A.\ Boette}
\affiliation{Departamento de F\'{\i}sica, FCE, Universidad Nacional de La Plata, C.C. 67, La Plata (1900), Argentina}
\affiliation{Instituto de F\'isica de La Plata, UNLP - CONICET, Argentina}

\author{C. Iemmi}
\affiliation{Departamento de F\'isica, FCEyN, Universidad de Buenos Aires, Pabell\'on 1, Ciudad Universitaria,  Buenos Aires (1428), Argentina}
\affiliation{Consejo Nacional de Investigaciones Cient\'ificas y T\'ecnicas (CONICET), Argentina.}

\author{R.\ Rossignoli}
\affiliation{Departamento de F\'{\i}sica, FCE, Universidad Nacional de La Plata, C.C. 67, La Plata (1900), Argentina}
\affiliation{Instituto de F\'isica de La Plata, UNLP - CONICET, Argentina}
\affiliation{Comisi\'on de Investigaciones Cient\'{\i}ficas de la Provincia de Buenos Aires (CIC), La Plata (1900), Argentina}

\author{S. Ledesma}
\affiliation{Departamento de F\'isica, FCEyN, Universidad de Buenos Aires, Pabell\'on 1, Ciudad Universitaria,  Buenos Aires (1428), Argentina}
\affiliation{Consejo Nacional de Investigaciones Cient\'ificas y T\'ecnicas (CONICET), Argentina.}

\begin{abstract}

We present an experimental optical implementation of a parallel-in-time discrete model of quantum evolution, based on the entanglement between the quantum system and a finite dimensional quantum clock. The setup is based on a programmable spatial light modulator which entangles the polarization and transverse spatial degrees of freedom of a single photon. 
It enables the  simulation of a qubit {\it history state} containing the whole evolution of the system, capturing its main features in a simple and configurable scheme. We experimentally determine the associated system-time entanglement, which is a measure of  distinguishable quantum evolution, and also the time average of observables, which in the present  realization can be obtained through one single measurement. 
\end{abstract} 
\maketitle

\section{Introduction}

Physics is a science that 
attempts to describe the behavior of natural systems, i.e., their evolution through time. 
In classical mechanics time is treated as an external classical parameter, assumption that remains in the standard formulation of quantum mechanics since 
probabilities are only assigned to  
observable measures made at a certain moment in time. In this sense, time  
reserves 
a special status in quantum mechanics.

The Newtonian notion of time, in which it is considered as a parameter essentially different from space coordinates,
was modified with the introduction of Lorentz transformations in relativity theory, but 
for each inertial frame it remains as a global external background parameter. In both cases, furthermore, it is assumed that the time coordinate can be read from an appropriate classical clock.
This assumption fails in quantum gravity, where the space-time metric is a dynamical object and must therefore be quantized, implying that a physical clock should be a quantum system itself \cite{Ro.04,Ku.11,IS.93,JT.11,Bo.11, Ho.12}.
Indeed, as predicted by the  Wheeler-DeWitt equation \cite{DW.67}, in quantum gravity ``there is no time''. Canonical quantization of general relativity preserves the constraint of a static state of the universe, and this lead essentially to the {\it problem of time}: the incompatibility between a timeless static description of the universe and the notion of time in the evolution of quantum systems. 

In the early 80's Page and Wootters proposed a mechanism \cite{PaW.83} to reconcile this apparent contradiction and since then the incorporation of time in a fully quantum framework has attracted increasing attention
\cite{Ga.09,GL.15,Ma.15,FC.13,FC.15,Er.17,Pa.17,Ni.18,Ll.18,BR.16,BR.18}. 
According to this {\it timeless approach} the universe is in a stationary state, and quantum evolution is explained by the entanglement between an evolving subsystem of the universe and a second quantum system, chosen as the reference clock. The ensuing {\it history state} contains the information about the whole evolution of the subsystem,  which can be recovered through appropriate measurements at the clock.  

 An experimental illustration  of these ideas 
 was presented in Ref.~\cite{Moreva14} using the polarization entangled state of two photons, one of which is used as  a two-dimensional clock to gauge the evolution of the second. More recently this realization has been extended to use the position of a photon, as a continuous variable, to describe time~\cite{Moreva17}. 
 
On the other hand, a fully discrete version of the formalism, based on a finite dimensional quantum clock, was developed  in 
 \cite{BR.16,BR.18}. Such scheme  leads to discrete history states,  which have the advantage that they  can be directly generated through a quantum circuit. Moreover, the associated Schmidt-decomposition and ensuing system-time entanglement can be easily obtained, with the latter representing a measure of distinguishable quantum evolution. 
 
 In the present work we introduce a simple optical implementation of such parallel-in-time {\it discrete} model of quantum evolution, in which the quantum clock has a finite configurable dimension $N$.
 This realization is carried out by using the polarization and the transverse spatial degrees of freedom (DOFs) of the light field to encode the emulated bipartite quantum system.
 Through the use of a programmable spatial light modulator ($\text{SLM}$) we generate non-separable states 
 sometimes called {\it classical entangled states}.
 The scheme enables the generation of discrete history states of a qubit, and hence to experimentally determine related quantities which characterize the quantum evolution, such as the associated system-time entanglement. Moreover, it allows us to recover time averages of observables of the system efficiently through  one single measurement, 
 instead of a set of $N$ sequential measurements. 
 
 The paper is organized as follows: We first provide, in Section \ref{sec:formalism}, a succinct description of the discrete formalism presented in Refs. \cite{BR.16,BR.18}. 
The experimental implementation and results are described in Section \ref{sec:impl},  where the modulation introduced by the SLM is analyzed in detail and expressed as unitary operators in polarization space. Theoretical and experimental results for time-averages are determined and compared. The 
ensuing system-clock entanglement is also analyzed for different trajectories,  and the so-called entangling power of the setup is as well discussed.  
Conclusions and perspectives are  finally presented in \ref{conclusions}.

\section{Formalism}\label{sec:formalism}

We consider a system $S$ and a reference clock system $T$ in a joint pure state $|\Psi\rangle \in \mathcal{H}_S\otimes \mathcal{H}_T$, with $\mathcal{H}_T$ of finite
dimension $N$. Any such state can be written as \cite{BR.16,BR.18}
\begin{eqnarray}\label{Eq:hist}
|\Psi\rangle=\frac{1}{\sqrt{N}}\sum_{t}|\psi_t\rangle|t\rangle
\label{1}
\end{eqnarray}
where ${\lbrace\vert t \rangle\rbrace}_{t=0}^{N-1}$ is an orthonormal basis of $T$ and $|\psi_t\rangle$ are states of $S$ (not necessarily orthogonal)  satisfying $\sum_t \langle \psi_t|\psi_t\rangle/N =
\langle\Psi|\Psi\rangle = 1
$. The state $|\Psi\rangle$ can describe, for instance, the whole
evolution of an initial pure state $|\psi_0\rangle$ of a physical system $S$ at a discrete
set of times, in which case $|\psi_t\rangle$ is the normalized state of the system at time $t$. Then, $|\psi_t\rangle$ can be recovered
as the conditional state of $S$ after a local measurement
at $T$ in the previous basis, with result $t$: If $\Pi_t=\mathbbm{1} \otimes |t\rangle\langle t|$, then 
\begin{eqnarray}
|\psi_t\rangle\langle \psi_t|=
\frac{{\rm Tr}_T \left(|\Psi\rangle\langle \Psi|\Pi_t\right)}{\langle \Psi|\Pi_t|\Psi\rangle}\,.
\end{eqnarray}
In shorthand notation, $|\psi_t\rangle=\sqrt{N}\langle t|\Psi\rangle$. Moreover, if $|\Psi\rangle$ is enforced to be an eigenstate of the unitary operator \cite{BR.18} 
\begin{equation}
{\cal U}=\sum_t U_{t,t-1}\otimes|t\rangle\langle t-1| \,,   
\end{equation}
where $U_{t,t-1}$ are arbitrary unitary operators satisfying the cyclic condition $U_{N,N-1}\ldots U_{1,0}=\mathbbm{1}_S$ (and $|t=N\rangle\equiv |t=0\rangle$),  then $|\psi_t\rangle$ follows a discrete unitary evolution \cite{BR.18}: $|\psi_t\rangle=U_t|\psi_0\rangle$ if ${\cal U}|\Psi\rangle=|\Psi\rangle$, with $U_t=U_{t,t-1}\ldots U_{1,0}$ (the eigenvalues of ${\cal U}$ are the $N$ $N^{\rm th}$ roots of unity, and $U_t\rightarrow e^{-i2\pi kt/N}U_t$, $k=1,\ldots,N-1$, for the other eigenvalues). Writing ${\cal U}=\exp[-i{\cal J}]$, the previous eigenvalue equation corresponds to ${\cal J}|\Psi\rangle=0$,  which is a generalized discrete version of the Wheeler-DeWitt equation \cite{BR.16}. 
And in the special case of a non-interacting ${\cal J}$, such that ${\cal J}=H_S\otimes \mathbbm{1}+\mathbbm{1}\otimes P_T$, 
then $U_t=\exp[-iH_St]$, with $H_S$ a Hamiltonian for system $S$ and $P_T$ a ``momentum'' for system $T$, both with eigenvalues $2\pi k/N$. Moreover, 
the equation ${\cal J}|\Psi\rangle=0$ then implies 
\begin{equation}
-\langle t|P_T|\Psi\rangle=H_S|\psi_t\rangle\,,
\end{equation}
which in the continuous limit obtained for large $N$ (and setting $\hbar=1$), 
reduces to the Schr\"odinger equation $i\partial_t|\psi_t\rangle=H_S|\psi_t\rangle$ \cite{BR.18}.  

The entanglement of the history state (\ref{1}) is a measure of the distinguishable evolution undergone by the system \cite{BR.16}.
If all states $|\psi_t\rangle$ are orthogonal, then $|\Psi\rangle$ is maximally entangled, whereas if all $|\psi_t\rangle$ are proportional (i.e., a stationary state), then $|\Psi\rangle$ becomes separable. Its entanglement entropy 
\begin{eqnarray}
    E(S,T)&=&\mathcal{S}(\rho_S)=\mathcal{S}(\rho_T)\,,\label{EST}\\
    \mathcal{S}(\rho)&=&-{\rm Tr}\,\rho\,\log_2\rho,\nonumber
\end{eqnarray}
where $\rho_{S,T}={\rm Tr}_{T,S}|\Psi\rangle\langle\Psi|=\sum_k\lambda_k |k_{S,T}\rangle\langle k_{S,T}|$ 
are the reduced system and clock states, respectively, 
then ranges from 0 for stationary states to $\log_2N$ when all $|\psi_t\rangle$ are mutually orthogonal. Thus, $2^{E(S,T)}$ is a measure of the number of distinguishable states visited by the system. Of course, 
when the system dimension $d_S$ is smaller than $N$, as will occur in the situation here considered,  the maximum number of orthogonal states $|\psi_t\rangle$ is $d_S$ and hence $E(S,T)\leq \log_2 d_S$. In general 
$E(S,T)\leq \log_2 M$, where $M\leq {\rm Min}[d_S,N]$ is the rank of $\rho_S$ or $\rho_T$ (identical). When $S$ is a qubit, $d_S=2$ and then $E(S,T)\leq \log_2 d_S=1$. 

We may also employ the quadratic entanglement
\begin{eqnarray}
    E^{(2)}(S,T)&=&\mathcal{S}_2(\rho_S)=\mathcal{S}_2(\rho_T)\\
    &=&\frac{2}{N}\left(N-1-\frac{2}{N}\sum_{t<t'}|\langle \psi_t|\psi_{t'}\rangle|^2\right)
\end{eqnarray}
where $\mathcal{S}_2(\rho)=2{\rm Tr}\,\rho(\mathbbm{1}-\rho)=2(1-{\rm Tr}\,\rho^2)$ is the quadratic entropy (also known as linear entropy, as it corresponds to $-\ln\rho\approx \mathbbm{1}-\rho$ in $S(\rho)$). This entropy can be directly evaluated without knowledge of the eigenvalues, and can be accessed experimentally through purity measurements of the reduced state $\rho_S$.  It is again a measure of the distinguishability between the evolved states. Its minimum value for an evolution between fixed initial and finial states due to a constant Hamiltonian $H_S$ is obtained for an evolution
 within the subspace generated by the initial and final
 states \cite{BR.18}, 
which proceeds precisely along the geodesic determined by the 
Fubini-Study metric \cite{AA.90, La.17,Ma.45,Bt.83}. 

While Eq.\ (\ref{EST}) is independent of the order of the states $|\psi_t\rangle$, it is also possible to consider the  entanglement entropies $E_{n}(S,T)$ associated with the first $n\leq N$ time-steps, determined by the partial history states $|\Psi_n\rangle=\frac{1}{\sqrt{n}}\sum_{t=0}^{n-1}|\psi_t\rangle|t\rangle$. 
Their variation with $n$  will provide information on the type of evolution. 
For instance, a periodic evolution will lead to an essentially  $n$-independent entanglement $E_{n}(S,T)$  (for a periodic evolution of period $L$, such that  $|\psi_{t+L}\rangle=e^{i\gamma_t}|\psi_{t}\rangle$ for $t=0,\ldots,L-1$, the entanglement  over $kL$ times is independent of the number of cycles $k$: 
$E_{kl}(S,T)=E_L(S,T_{L})$ \cite{BR.16}), while a steadily increasing $E_{n}(S,T)$ indicates increasing distinguishability of the visited  states.

\section{\label{sec:impl} Experimental implementation}

To provide an experimental realization of the concepts here discussed, we propose a full-optical architecture to generate the discrete history states of Eq.~(\ref{1}). We use the linear transverse momentum-position of single photons to set the time $|t\rangle$ of the quantum clock system $T$ and its polarization to encode the state $|\psi_t\rangle$ of the quantum system $S$. 
It should be noted that by encoding the subsystems in two different DOFs of a single particle \cite{PhysRevA.57.R1477,fiorentino2004,ndagano2017,kagalwala2017,imany2018}, 
the resulting non-separable state is not, strictly speaking, a nonlocal quantum entangled state: although such encoding is often referred to as ``entanglement" between DOFs, it has a local nature, while  
``true" quantum entanglement  occurs between  different particles~\cite{Karimi1172,Aiello_2015}.

\begin{figure}[htp]
\centering
\includegraphics[width=0.43\textwidth]{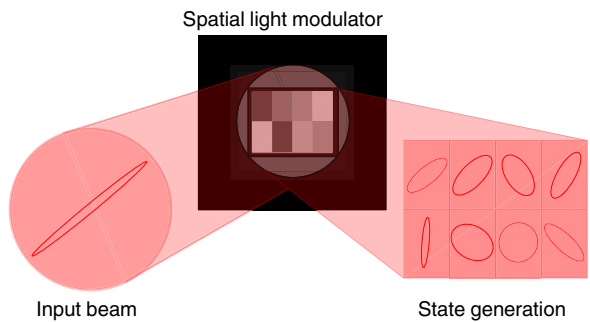}
\caption{Sketch of the action of the
SLM on the initial state of light. The complex function $\Gamma(\mathbf{x})$ programmed on the SLM defines eight spatial regions in the ROI, and each of them modifies the polarization state according to the particular gray level.} 
\label{fig:schematic-codification}
\end{figure}

One of the simplest ways to accomplish this encoding is to use a programmable SLM as a means to create correlations between polarization
and spatial DOF of photons \cite{nagali2010,fickler2014,lemos2014}. In general, this kind of devices allows to coherently modulate the amplitude, phase and polarization of the electromagnetic field. 
It is thus possible to display different regions on the SLM screen and vary, in each of these regions, the polarization of the light field keeping constant its amplitude and phase. It leads to a state generation scheme as that indicated in Fig.~\ref{fig:schematic-codification} where, as an example, eight independent rectangular regions are addressed on the SLM, each one with a different constant function modulation.

\subsection{\label{sec:state_generation} Generation of discrete history states}

The history state $|\Psi\rangle$ in Eq.~(\ref{Eq:hist}) can be generated from an initial product state $|\psi_0\rangle|0\rangle$ as \begin{equation}\label{Eq:histGen}
    |\Psi\rangle=\mathcal{W}(\mathbbm{1}\otimes H)|\psi_0\rangle|0\rangle,
\end{equation}
where $H$ is a Hadamard-like gate on
the clock system $\big(H|0\rangle= \frac{1}{\sqrt{N}} \sum_{t=0}^{N-1}|t\rangle \big)$ and 
$\mathcal{W}=\sum_{t=0}^{N-1} U_t{\otimes }|t\rangle\langle t|$
is the control-$U_t$ gate. For $t=0,1,...,N-1$, the $U_t~'s$ are unitary operators on the system $S$, so that
$|\psi_t\rangle=U_t|\psi_0\rangle$. 

In our experimental implementation, the initial state is a photonic state defined by the product of its polarization state ($|\psi_0\rangle$), 
and its spatial state $(|0\rangle)$ described by the transverse wavefront profile.
By using a formalism similar to that of previous works \cite{solis2013,lemos2014}
where the polarization or the transverse spatial DOF of photons are manipulated through the use of SLMs, the generation of history states can be explained as follows:

A paraxial and
monochromatic single-photon field, assumed here to be in a pure state, is described by 
\begin{equation}
|\psi_0\rangle|0\rangle=\sum_{\mu}\int\!d\mathbf{x}\,\alpha_\mu f(\mathbf{x})|\mu\rangle|1\mathbf{x}\rangle,
\end{equation}
where $\mu$
runs over two orthogonal polarizations, $\mathbf{x}=(x,y)$ is the transverse position coordinate, and $f(\mathbf{x})$ is the normalized transverse probability
amplitude for this state, i.e., $\int\!d\mathbf{x}\,|f(\mathbf{x})|^2=1$. The SLM introduces a
polarization-dependent modulation that
can be ideally interpreted as the action of the operator 
\begin{equation}
    \Gamma=\sum_{\alpha, \beta}\int d\mathbf{x}\,\Gamma_{\alpha \beta}(\mathbf{x})|\alpha,1 \mathbf{x}\rangle\langle\beta, 1\mathbf{x}|,
\end{equation}
so that, after impinging the SLM, the state of the photon field reads
\begin{equation}
|\Psi\rangle\propto\sum_{\mu,\nu}\label{Eq:transf} \int\!d\mathbf{x}\,\alpha_\nu f(\mathbf{x})\Gamma_{\mu\nu}(\mathbf{x})|\mu\rangle|1\mathbf{x}\rangle.
\end{equation}

Let us consider a modulation distribution $\Gamma_{\mu\nu}(\mathbf{x})$ defining an array of $N$ rectangular and adjacent spatial regions of width $2a$, and length $2b$. On each of these regions we have a constant complex modulation, $C_{\mu\nu}^{(t)}$. Thus,

\begin{equation} \label{eq:fNregions}
\begin{split}    
\Gamma_{\mu\nu}(\mathbf{x})&= \sum_{t=0}^{N-1}C_{\mu\nu}^{(t)}~{\rm
rect}\!\left(\frac{x-x_t}{2a}\right){\rm
rect}\!\left(\frac{y-y_t}{2b}\right)\\
&C_{\mu\nu}^{(t)}=c_{\mu\nu}^{(t)}{\rm e}^{i\gamma^{(t)}_{\mu\nu}},~~c_{\mu\nu}^{(t)}\geq0
\end{split}
\end{equation}
where ${\rm
rect}\!\left(u\right)=1$ if $|u| < \frac{1}{2}$, 0 in other case, and  
the centres of these regions are in $\{(x_t,y_t)\}_{t=0}^{N-1}$, with $x_t=a, 3a,5a,...,$ and $y_t=b, 3b, 5b,...$\,. With this prescription we can define the spatial states 
\begin{equation}
|t\rangle=\frac{1}{\sqrt{\mathcal{N}_t}}\int d\mathbf{x}\,{\rm rect}\!\left(\frac{x-x_t}{2a}\right){\rm rect}\!\left(\frac{y-y_t}{2b}\right)|1\mathbf{x}\rangle,
\end{equation}   
which form an orthonormal basis of the discretized spatial Hilbert space 
of the single photon. Finally, by combining this result with Eq.~(\ref{eq:fNregions}), the transformed state in Eq.~(\ref{Eq:transf}) can be written in the following way: 
\begin{equation}
|\Psi\rangle\propto\sum_{t=0}^{N-1}\sum_{\mu,\nu}\alpha_\nu C_{\mu\nu}^{(t)}|\mu\rangle|t\rangle.\label{Psi_mod} 
\end{equation}

In our implementation the modulation introduced by the SLM implies a transformation  only of the polarization DOF.  It means that $\sum_{\mu\nu}\alpha_\nu C_{\mu\nu}^{(t)}|\mu\rangle|t\rangle= (U_t\otimes\mathbbm{1})|\psi_0\rangle|t\rangle\equiv |\psi_t\rangle|t\rangle$, with $U_t$ a unitary operator and $|\psi_t\rangle$ the polarization state associated to the $t$-spatial region. Therefore, the SLM transforms the initial photon state as
\begin{equation}
\begin{split}
|\psi_0\rangle|0\rangle\stackrel{\rm{SLM}}{\Longrightarrow}~& \frac{1}{\sqrt{N}}\sum_{t=0}^{N-1}U_t|\psi_0\rangle|t\rangle={\cal W}\left(|\psi_0\rangle\sum_{t=0}^{N-1}\frac{1}{\sqrt{N}}|t\rangle\right)\\
&={\cal W}\left(\mathbbm{1}\otimes H\right)|\psi_0\rangle|0\rangle,
\end{split}
\label{eq:Uslm}
\end{equation}
and thus generates the history state as expressed in Eq.~(\ref{Eq:histGen}), where the system $S$ and the clock system $T$ are emulated by the polarization and spatial DOFs, respectively.

\subsection{Setup and measurements}

\begin{figure}[htp]
\centering
\includegraphics[width=0.43\textwidth]{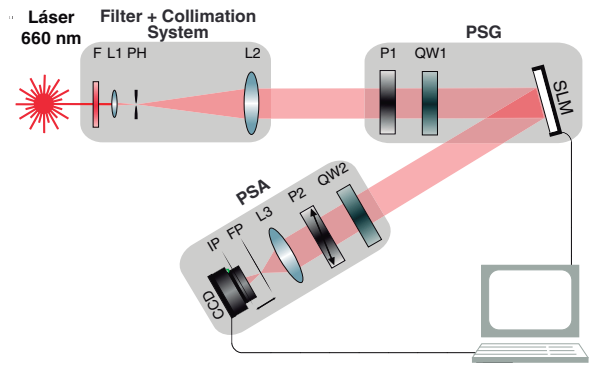}
\caption{Experimental setup used for history-states generation and subsequent characterization of the evolution of the quantum system $S$.} 
\label{fig:setup}
\end{figure}

The experimental setup used for simulating the parallel-in-time quantum evolution is sketched in Fig.~\ref{fig:setup}. In the first part, a $660\text{nm}$ solid state laser beam is expanded, filtered and collimated in order to illuminate a $\text{SLM}$ with a planar wave with approximately uniform amplitude distribution over the region of interest (ROI). 
This SLM, based on a reflective liquid crystal-on-silicon (LCoS) micro-display, with a spatial resolution of 1024x768 pixels, is used to represent the whole system $|\Psi\rangle$ of Eq.~(\ref{Eq:hist}). It gives the possibility to dynamically address the optical function on the screen, pixel by pixel.
In particular, the SLM used in our experiment, consists of a HoloEye Lc-R 2500 in combination with a polarizer (P1) and a quarter wave plate (QW1) that provide the adequate incoming state of light to obtain the maximum range of polarization modulation. 
This is obtained from a Mueller-Stokes characterization of the LCoS \cite{marquez2001,marquez2008},  followed by an optimization to have a wide range of {\it pure polarization modulation}, i.e., 
without any additional global phase due to an optical path difference, regardless of the gray level that the pixels of the LCoS are set for.
Therefore, as each pixel is controlled individually, we can program a particular function $\Gamma(\mathbf{x})$ 
which characterizes the modulation distribution. Then, the wavefront of the electromagnetic field acquires a specific polarization conditioned on the transverse position in the plane of
the SLM. 

In the second part of the setup, a polarization state analyzer (PSA) is used for the initial characterization of the SLM as a polarization state generation (PSG).
For this purpose, after reflection on the
SLM, the outgoing beam is focused by the lens L3 onto the detection plane, which is chosen to match the image plane (IP) or the Fourier plane (FP). A quarter wave plate (QW2) and a linear polarizer (P2) project the polarization state of the light beam in the different states of the reconstruction basis. Intensity measurements are recorded in the IP or in the FP, depending on the characterization for amplitude or phase modulation, respectively.

In addition, and as a proof-of-principle demonstration, we
have inserted neutral-density filters, previous to the PSG stage, to highly attenuate the power of the laser beam at the single-photon regime in such a way that it corresponds to the presence of less than one photon, on average, at any time, in the experiment.
This pseudo single-photon source can be used to
mimic a single-photon state, and 
as is usual in optical implementations of quantum simulations or quantum-states estimation \cite{Malik2014,QPS17,Lima2018}, it is enough to test the feasibility of the proposed method for simulating the main features of a parallel-in-time quantum evolution. Besides, instead a CCD camera, we used a high sensitive camera based on CMOS
technology (Andor Zyla 4.2 sCMOS) to carry out the intensity measurements in this regime.

\begin{figure}[htp]
\centering
\includegraphics[width=0.47\textwidth]{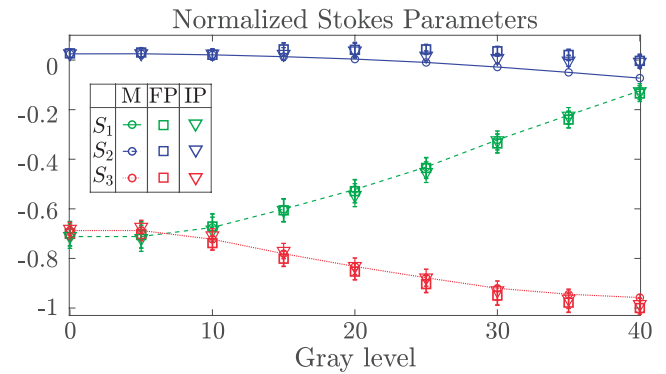}
\caption{Stokes parameters of the polarization state reflected by the SLM as a function of the gray level. The incoming polarization state is given by $(S_0,S_1,S_2,S_3)=(1.000,0.040,0.951,-0.026)$. The graphic shows the experimental values obtained by a measurement in the IP ($\bigtriangledown$) or in the FP ($\Box$), in comparison with those predicted by the Mueller matrix ($\bigcirc$).}
\label{fig:stokes}
\end{figure}
\begin{table*}[htp]
  \begin{center}
    \begin{tabular}{ccrr}
    \hline
 & {\rm Trajectory} &\multicolumn{1}{c}{$\;t_1(0)$} &\multicolumn{1}{c}{$\;\;\;t_2(30)$} \\
     \hline
 {$\langle\sigma_{x}
\rangle$} &  &-0.7328&\; -0.3183   \\
 $\langle\sigma_{y}
\rangle$ & N$^o$ 1 &-0.6621 &\;-0.9365   \\
 $\langle\sigma_{z}\rangle$ && 0.0541 &\;0.0385  \\
    \hline
 \end{tabular}
 \vspace*{0.25cm}

    \begin{tabular}{ccrrrr|crrrr}
    \hline
    & {\rm Trajectory} &\multicolumn{1}{c}{$\;\;t_1(0)$} & \multicolumn{1}{c}{$\;\;\;t_2(30)$} & \multicolumn{1}{c}{$\;\;\;t_3(0)$} & \multicolumn{1}{c|}{$\;\;\;t_4(30)$} &$\;$ {\rm Trajectory} &\multicolumn{1}{c}{$\;\;\;t_1(25)$} & \multicolumn{1}{c}{$\;\;\;t_2(0)$} & \multicolumn{1}{c}{$\;\;\;t_3(15)$} & \multicolumn{1}{c}{$\;\;\;t_4(35)$}\\
    \hline
{$\langle\sigma_{x}
\rangle$} &  & -0.7358&\; -0.3465&\; -0.7122 & \;-0.3147\;  &   & -0.4558 &\; -0.7218 &\; -0.6019 &\; -0.2006   \\
 $\langle\sigma_{y}
\rangle$ & N$^o$ 2
&-0.6505 &-0.9306  & -0.6802 & -0.9299\; & $\;$N$^o$ 5
&-0.8689 &-0.6673  & -0.7789 & -0.9632     \\
 $\langle\sigma_{z}\rangle$ & &0.0447 & 0.0273 & 0.0452 & 0.0271\;
    & ~ & 0.0394& 0.0438& 0.0475 & 0.0134\\
    \hline
    & {\rm Trajectory} &\multicolumn{1}{c}{$\;\;t_1(0)$} & \multicolumn{1}{c}{$\;\;\;t_2(30)$} & \multicolumn{1}{c}{$\;\;\;t_3(30)$} & \multicolumn{1}{c|}{$\;\;\;t_4(0)$} &$\;$ {\rm Trajectory} &\multicolumn{1}{c}{$\;\;\;t_1(0)$} & \multicolumn{1}{c}{$\;\;\;t_2(15)$} & \multicolumn{1}{c}{$\;\;\;t_3(25)$} & \multicolumn{1}{c}{$\;\;\;t_4(35)$}\\
    \hline
{$\langle\sigma_{x}
\rangle$} & &-0.7093 &\; -0.2996 &\; -0.3614 &\; -0.7277\; &  & -0.7312 &\; -0.6167 &\; -0.4331 & -0.2030  \\
 $\langle\sigma_{y}
\rangle$ & N$^o$ 3 
&-0.6787 &-0.9404  & -0.9214 & -0.6551\; & $\;$N$^o$ 6 
& -0.6491 &-0.7692  & -0.8889 & -0.9627  \\
 $\langle\sigma_{z}\rangle$ & &0.0453 & 0.0253 & 0.0365 & 0.0442\; & & 0.0420& 0.0433& 0.0413 & 0.0120 \\
     \hline
    \end{tabular}
    \vspace*{0.25cm}

    \begin{tabular}{ccrrrrrrrr}
    \hline
  & {\rm Trajectory} &
\multicolumn{1}{c}{$t_1(0)$} & \multicolumn{1}{c}{$\;\;t_2(30)$} & 
\multicolumn{1}{c}{$\;\;t_3(0)$} & \multicolumn{1}{c}{$\;\;t_4(30)$} &\multicolumn{1}{c}{$\;\;t_5(0)$} &\multicolumn{1}{c}{$\;\;t_6(30)$} &\multicolumn{1}{c}{$\;\;t_7(0)$} &\multicolumn{1}{c}{$\;\;t_8(30)$}\\
    \hline
 {$\langle\sigma_{x}
\rangle$} &  &-0.7110&\; -0.3183 &\; -0.6849 &\; -0.2957&\;  -0.7112&\; -0.3432&\; -0.7315&\; -0.3496 \\
 $\langle\sigma_{y}
\rangle$ & N$^o$ 4
&-0.6614 &-0.9382  & -0.7017 & -0.9303 & -0.6475 & -0.9263 & -0.6473 &\;-0.9180 \\
 $\langle\sigma_{z}\rangle$ && 0.0382 & 0.0288 & 0.0433 & 0.0183 & 0.0334 &  0.0322 &  0.0376 & 0.0337   \\
    \hline
    & {\rm Trajectory} &
\multicolumn{1}{c}{$t_1(0)$} & \multicolumn{1}{c}{$\;\;t_2(10)$} & 
\multicolumn{1}{c}{$\;\;t_3(15)$} & \multicolumn{1}{c}{$\;\;t_4(20)$} &\multicolumn{1}{c}{$\;\;t_5(25)$} &\multicolumn{1}{c}{$\;\;t_6(30)$} &\multicolumn{1}{c}{$\;\;t_7(35)$} &\multicolumn{1}{c}{$\;\;t_8(40)$}\\
    \hline
 {$\langle\sigma_{x}
\rangle$} && -0.7538 &-0.6908&-0.5848 &-0.5458 &-0.4772& -0.3308 & -0.2489&-0.0965   \\
 $\langle\sigma_{y}
\rangle$ & N$^o$ 7&-0.6383&-0.6910& -0.8004& -0.8194&-0.8705 &-0.9408  &-0.9656&-0.9867    \\
 $\langle\sigma_{z}\rangle$ & & 0.0595&0.0473 &0.0548  &0.0495 &0.0447&0.0387 &   0.0315&0.0009  \\
 \hline
 \end{tabular}
  \end{center}
  
 \caption{Generated history states corresponding to two time-steps (trajectory 1), four time-steps (trajectories 2, 3, 5 and 6) and eight time-steps (trajectories 4 and 7).  The discrete evolution of the system is seen as a trajectory on the Bloch sphere. The initial state of the system $|\psi_0\rangle$ is described by the Stokes vector $(S_0,S_1,S_2,S_3)=(1.000,0.040,0.951,-0.026)$. In parentheses, the gray levels used to experimentally implement each trajectory.}
 \label{table:steps}
 \end{table*}

In Fig.~\ref{fig:stokes} we plot the Stokes parameters of the state prepared by the SLM when a single gray level, between 0 and 40, is addressed on the whole screen. In any case the polarization of the input state is $(S_0,S_1,S_2,S_3)=(1.000,0.040,0.951,-0.026)$. The graphic shows the parameter values obtained as a measurement in the IP ($\bigtriangledown$) or in the FP ($\Box$), in comparison with those predicted by the Mueller matrix ($\bigcirc$). For these range of gray levels, all the values are in good agreement
which indicates a good performance of the whole setup for the modulation of the polarization state and subsequent characterization of such states.
We should mention that, while it is possible to set gray levels up to 255, for those above 40 
the depolarization due to
temporal phase fluctuations of the employed SLM becomes important. In fact,
devices based on LCoS technology may lead to a flicker in the optical beam because of the
digital addressing scheme (pulse width modulation) which
introduces, among other undesirable effects, those phase
fluctuations \cite{lizana2008,lizana2010} that affect the quality of the state that is intended to encode. 

Once the modulation of the SLM was fully characterized, the same PSG-PSA system was used for experimentally perform the system-time history state 
$|\Psi\rangle$, 
and the subsequent characterization of the discrete unitary evolution of the system state $|\psi_t\rangle=U_{t}|\psi_0\rangle$
($t=0,...,N-1$). For gray levels between 0 and 40, different history states $|\Psi\rangle$ were generated with 2, 4, and 8 time steps.  These history states are displayed in Table \ref{table:steps}. 
According to our experimental implementation, each state $|\psi_t\rangle$ visited by the system is specified in terms of the mean values $\langle\sigma_\mu\rangle$ of the Pauli operators $\sigma_\mu$, which are just the measured Stokes parameters: $S_{1}=\langle{\sigma}_{z}\rangle$, $S_{2}=\langle{\sigma}_{x}\rangle$, and $S_{3}=\langle{\sigma}_{y}\rangle$ (see subsection \ref{sec:Evolution operators}). Trajectories 1--4 employ gray levels  $0$ and $30$, 
trajectories 5--6 gray levels 0, 15, 25 and 35, while  7 uses gray levels 0, 10, 15, 20, 25, 30, 35 and 40. 
These trajectories were chosen in order to compare, for example, ``equivalent'' (e.g., trajectories 2 and 3 or trajectories 
5 and 6, which differ just in the order of gray levels) 
 and ``non-equivalent'' (e.g.\ trajectories 
 4 and  7) sets of gray levels, or to compare between essentially periodic (trajectories  2 and  4) and non-periodic (trajectories 
 1 and  3) evolutions.

\subsection{\label{sec:Polarization measurement} 
System evolution and mean values}

In the previous subsection we have described how our setup generates history states within a parallel-in-time discrete model of quantum evolution. 
This implementation allows us to compute
the time-average of system observables 
throughout its evolution in two different ways: 
\begin{itemize}
    \item From the set of measurements which are performed, sequentially, on the system $S$.
    \item From a  single measurement that involves information of the whole evolution of the system $S$. 
\end{itemize}
In fact, let us consider an operator 
$A={O}\otimes \mathbbm{1}$, with $\mathbbm{1}$ the identity operator on the clock system and $O$
an observable of the system $S$. Then, its  expectation value in the full history state $|\Psi\rangle$ is given by 
\begin{eqnarray}
\langle A\rangle_\Psi& \equiv &\langle\Psi\vert O \otimes \mathbbm{1}\vert \Psi \rangle\nonumber\\
&=&\frac{1}{N}\sum_{t=0}^{N-1}\langle \psi_t\vert O \vert \psi_t \rangle\,,\label{Eq:av}
\end{eqnarray}
which represents the time-average $\overline{\langle O \rangle}=\frac{1}{N}\mathrm{Tr}_S(\sum_{t=0}^{N-1}O \vert \psi_t \rangle\langle \psi_t \vert)$.

In our experimental scheme, we can identify the observable $O$ with one of the Pauli operators $\sigma_{\mu}$. In order to test these two approaches we perform a
proper polarization measurement to 
compute the time-average $\overline{\langle \sigma_{\mu} \rangle}$ for different evolutions of the system $S$. For this purpose the PSA is used to project the polarization state of the incoming beam and record the intensity of the non-extinguished beam.
On one hand, if an intensity measurement is performed in the IP, the mean values of the Pauli operators $\sigma_\mu$, will vary from one of the spatial regions defined in Eq.~(\ref{eq:fNregions}) to the other, depending on the modulation $C_{\mu\nu}^{(t)}$ assigned to each of these regions. If the polarization state associated to the region $t$ is $|\psi_t\rangle$, $\sigma_\mu$ will have the mean value $\langle \psi_t|\sigma_\mu|\psi_t\rangle$ on this region, and the average on the full ROI is then computed as $\frac{1}{N}\sum_{t=0}^{N-1} \langle \psi_t|\sigma_\mu|\psi_t\rangle$. 
On the other hand, if an intensity measurement is performed in the FP, each of the spatial regions addressed on the SLM contribute to build the interference pattern. However, it is not possible to relate a spatial region in the IP to a particular region in the FP. 
The mean values are then given by $\langle \Psi|\sigma_\mu\otimes\mathbbm{1}|\Psi\rangle$, which implies a global measure in the FP. 
These two quantities are of course the same, since in the absence of optical losses, the total intensity of the non-extinguished beam is involved in their calculation, as expressed in Eq.\ (\ref{Eq:av}).  

\begin{figure}[htp]
\centering
\includegraphics[width=0.470\textwidth]{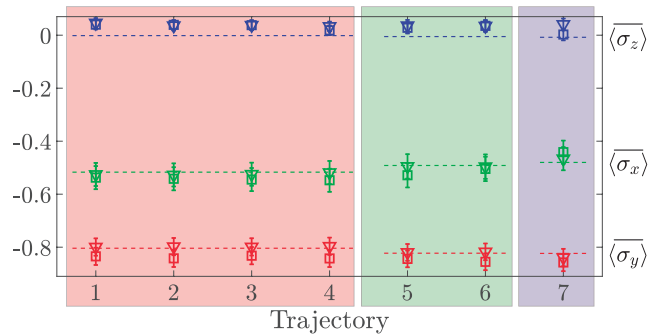}
\caption{Time averages of the polarization observables, $\overline{\langle\sigma_{\mu}
\rangle}$, for different evolutions of the initial system state $|\psi_0\rangle$.  The graphic shows the predicted values ($--$) in comparison with the experimental values, obtained by averaging over all spatial regions ($\Box$), and by means of one single measurement ($\bigtriangledown$). 
Each sector, identified by a single color, indicates   different evolutions through the same states $|\psi_t\rangle$.} 
\label{mean-values}
\end{figure}

Therefore, if we think of $|\Psi\rangle$ as a history state, our scheme provides an efficient method for the evaluation of the time-averaged polarization of the system throughout its trajectory. In fact, results shown on Fig.~\ref{mean-values} exhibit an excellent agreement between both experimental measurements, and between these and the theoretical values. 
In this plot we can see the time averages $\overline{\langle\sigma_{\mu}
\rangle}$ of the history states described in Table \ref{table:steps}, which correspond to different evolutions of the same initial state $|\psi_0\rangle$. 
As expected, these time averages have all the same values for trajectories 1--4, and for trajectories 5--6. 


\begin{figure}[htp]
\hspace*{-.20cm}\includegraphics[width=.5\textwidth]{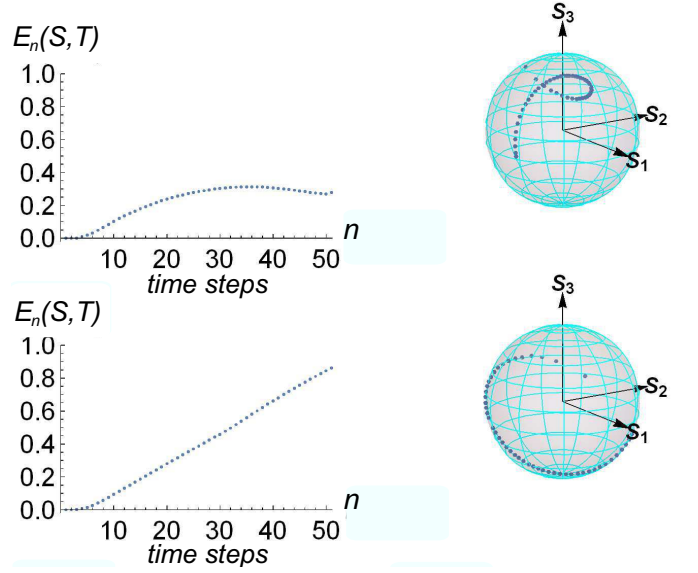}

\caption{System-time entanglement entropy $E_n(S,T)$ vs.\ number of steps (left panels) and 
 trajectories in the polarization Bloch sphere (right panels) for two different initial states. The  same set of 52 unitary evolution operators extracted from the experimental characterization was employed in both cases.  \ref{table:steps}} 
\label{figest5}
\end{figure}

In subsection \ref{sec:state_generation} we stated that the modulation introduced by the SLM can be described by a unitary transformation in polarization space. Experimentally, the modulation associated to a given gray level on the screen is described by a $4\times 4$ Mueller matrix $M$ \cite{marquez2008}. The Mueller matrix acts as a linear transformation on the polarization state of the light field represented by the Stokes vector $\bm S$, defined as
\begin{equation}
\bm S=\begin{pmatrix}S_0\\  S_1\\ S_2\\ S_3\end{pmatrix}=\begin{pmatrix}P_{00}+P_{\frac{\pi}{2}0}\\  P_{00}-P_{\frac{\pi}{2}0}\\ P_{\frac{\pi}{4}0}+P_{-\frac{\pi}{4}0}\\ P_{\frac{\pi}{4}\frac{\pi}{2}}+P_{-\frac{\pi}{4}\frac{\pi}{2}}\end{pmatrix},
\end{equation}
where the vector coefficients $P_{\theta\phi}$ are the results of six polarization measurements: horizontal and vertical linear polarization $(P_{00},P_{\frac{\pi}{2}0})$, $+45$ and $-45$ linear polarization $(P_{\frac{\pi}{4}0}, P_{-\frac{\pi}{4}0})$, and right and left circular polarization $(P_{\frac{\pi}{4}\frac{\pi}{2}}, P_{-\frac{\pi}{4}\frac{\pi}{2}})$. Within the quantum formalism, such measurements correspond to projections onto the polarization states $|P_{\theta\phi}\rangle=\cos(\theta)|H\rangle+e^{i\phi}\sin(\theta)|V\rangle$, so that we have  $S_{1,2,3}=\langle\sigma_{z,x,y}\rangle\,$, 
provided $P_{00}+P_{\frac{\pi}{2}0}=1$ and
$\sigma_{\mu}~'s$ are the Pauli operators defined with respect to the basis $\{|H\rangle, |V\rangle\}$. The polarization state of a single photon is therefore given by
\begin{equation}
    \rho=\frac{1}{2}\left(I+\bm r\cdot\bm\sigma\right),
\end{equation}
with $\bm r=\frac{1}{S_0}(S_1,S_2,S_3)$, and a unitary transformation in polarization space corresponds then to a rotation of the Bloch vector $\bm r$, which will be associated to a Mueller matrix of the form
\begin{equation}
M_R=\begin{pmatrix}
    1&&\bm 0\\
    \bm 0&& m_R 
    \end{pmatrix},\label{ret}
\end{equation}
where $m_R$ denotes an arbitrary $3\times 3$ rotation matrix. A Mueller matrix such as that describes the effect of an ideal retarder. However, the SLM used in our implementation introduces not only retardance but also diattenuation. Therefore the Mueller matrix associated to a given gray level will not have the form (\ref{ret}) that maps to a unitary transformation in polarization space. It is possible, nonetheless, to extract from a general Mueller matrix a pure retardance matrix that accounts for the effective phase transformation introduced by the optical system, by means of the Lu-Chipman decomposition \cite{Lu.96}. In this way, from the Mueller matrices obtained from the experimental characterization of the SLM we extracted a set of unitary matrices that describe the transformations performed on the polarization of the photon field, for 52 gray levels between 0 and 255, in steps of 5. 
The set of unitary matrices described above allows us to simulate history states
beyond those that we have actually implemented. The right panels in Fig.\ 
\ref{figest5} show examples of such simulated evolutions of the photon polarization as trajectories on the Bloch sphere.

In Fig.~\ref{figest5}, we also depict in the left panels the entanglement entropies $E_{n}(S,T)$ associated with the trajectories determined by the $52$ unitaries derived from the $N=52$ experimentally determined Mueller matrices, for two different initial states, as a function of the number $n$ of steps. For $n=N$, $E_{n}(S,T)$ becomes the system-time entanglement entropy $E(S,T)$ of the full trajectory. In the top and central panels time-ordering corresponds to increasing gray levels. In the top panel the 
trajectory exhibits a loop starting at step $n\approx 35$, implying a decreasing distinguishability between evolved states in this sector, which is reflected in a decrease of $E_{n}(S,T)$  for $n\agt 35$.  In contrast, in the central panel $E_{n}(S,T)$ increases linearly with $n$ as the trajectory has no loops and does not cross itself. 

\begin{figure}[htp]
\hspace*{-.20cm}\includegraphics[width=.5\textwidth]{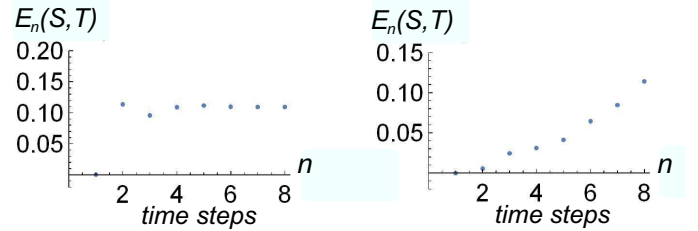}

\caption{System-time entanglement entropy $E_n(S,T)$ vs.\ number of steps, for 
the experimentally generated history states corresponding to the two eight time-steps trajectories of Table \ref{table:steps}.}
\label{figest6}
\end{figure}

Figure (\ref{figest6})  depict the entropies $E_{n}(S,T)$ for the two experimental eight-step trajectories of Table \ref{table:steps}.
 That on the left stays approximately constant after the third step, since it is determined  by a configuration with just two gray levels and the trajectory essentially oscillates between two  non-orthogonal states $|\psi_0\rangle$ and $|\psi_1\rangle$. 
In this case  $|\Psi_n\rangle=\frac{1}{\sqrt{n}}[
|\psi_0\rangle(\sum_{t\,{\rm even}}^{n-1}|t\rangle)+|\psi_1\rangle(\sum_{t\,{\rm odd}}^{n-1}|t\rangle)]$ and the  exact theoretical value of  $E_{n}(S,T)$ is given by 
\begin{equation}
E_{n}(S,T)=-\sum_{\nu=\pm}p_{n,\nu}\log_2 p_{n,\nu}\,,
\end{equation}
where the probabilities $p_{n,\pm}$ are $n$-independent for $n$ even while for $n$ odd they  
rapidly approach the same  even values as $n$ increases:
\begin{equation}p_{n,\pm}=\left\{\begin{array}{lr}
\frac{1}{2}[1\pm
|\langle \psi_0|\psi_1\rangle|]\,,&n\;{\rm even}\\
\frac{1}{2}\left[1\pm\sqrt{|\langle\psi_0|\psi_1\rangle|^2
(1-\frac{1}{n^2})+\frac{1}{n^2}}\right]\,,&n\;{\rm odd}\end{array}\right.\,.\end{equation}
The observed value  $E_{n}(S,T)\approx 0.11$ (for $n$ even or $n\agt 5$ if odd) is then in agreement with the overlap $|\langle \psi_0|\psi_1\rangle|\approx 0.97$ between both states. This almost  periodic trajectory is compatible with an approximately constant effective Hamiltonian 
$H=\frac{\pi}{2}\bm{n}\cdot\bm{\sigma}$, where $\bm{n}$ is a vector in the plane spanned by the Bloch vectors of $|\psi_0\rangle$ and $|\psi_1\rangle$, halfway between both states, such that $e^{-iH}$ is a rotation of angle $\pi$ around this axis and  $e^{-iH}|\psi_0\rangle=|\psi_1\rangle$, $e^{-iH}|\psi_1\rangle=|\psi_0\rangle$. 

On the other hand, on the bottom right panel, $E_{n}(S,T)$ increases almost linearly for $t\agt 5$, 
reflecting a trajectory where the distinguishability between the evolved state $|\psi_t\rangle$ and the initial state increases monotonically. In this case the evolved states lie approximately within a plane and  the trajectory is approximately compatible with a Hamiltonian $H=\alpha_t\bm{n}\cdot\bm{\sigma}$, with $\bm{n}$ orthogonal to this plane and varying strength $\alpha_t$ (or equivalently, constant $\alpha_t$ and varying time intervals). 

We  mention that  the behavior of the quadratic  entropy $E^{(2)}_{n}(S,T)$ is completely similar to that of $E_{n}(S,T)$,  since the polarization reduced state $\rho_S$ is a qubit state. And for a qubit, $S_2(\rho_S)$ is just an increasing (and concave) function of the von Neumann entropy $S(\rho_S)$. 

\subsection{\label{sec:Evolution operators} Evolution operators and entangling power}

For any of these simulated history states we can now reconsider the generating operator ${\cal W}=\sum_t U_t\otimes |t\rangle\langle t|$, which can be here expressed as 
\begin{equation}
{\cal W}=\sum_t \left(\frac{1}{2} {\sum}'_\mu r_{\mu}(t) \sigma_{\mu}\right) \otimes |t\rangle\langle t|={\sum}'_{\mu} \lambda_{\mu}\tilde{\sigma}_{\mu}\otimes O_{\mu}.
\end{equation}
Here we have first expanded the unitary operators in polarization space in the Pauli operators plus $\sigma_0=\mathbbm{1}$, with $r_\mu(t)={\rm Tr}\,U_t\sigma_\mu$ (and $\sum'_\mu=\sum_{\mu=0}^3$), and then written the ensuing Schmidt decomposition \cite{BR.18}, where 
$\tilde{\sigma}_{\mu}$ and $O_{\mu}$ are orthogonal operators in polarization and spatial spaces (${\rm Tr}\,(\tilde{\sigma}_{\mu}^{\dagger}\tilde{\sigma}_{\nu})=2\delta_{\mu\nu}$, ${\rm Tr}(O_{\mu}^{\dagger}O_{\nu})= N\delta_{\mu\nu}$). The real non-negative numbers $\lambda_{\mu}$ are the Schmidt coefficients, which are the singular values of the  $4\times N$  matrix $C_{\mu,t}= r_{\mu}(t)/\sqrt{N}$ and satisfy $\sum_{\mu}\lambda_{\mu}^2=1$.    

Its quadratic operator entanglement \cite{BR.18}, $E^{(2)}({\cal W}) = 2(1-\sum_{\mu}\lambda_{\mu}^4)$, which depends on the unitary evolution operators $U_t$ but not on the initial state, is  proportional to the {\it entangling power} of ${\cal W}$ \cite{BR.18}, which is the average quadratic entanglement it generates when applied to 
  initial product states $|\psi_0\rangle H^{\otimes n}|0\rangle$ ($N=2^n$):  
\begin{equation}
 \langle E^{(2)} (S,T)\rangle=\frac{d_S}{d_S+1}E^{(2)}({\cal W})\,, \label{rel}
\end{equation}
where $d_S$ is the dimension of the system ($d_S=2$ in the present case) and 
 \begin{equation}
 \langle E^{(2)}(S,T)\rangle=
 \int_{\cal H} 2(1- {\rm Tr}\,\rho_S^2) d \psi_0\, \label{ME2}
\end{equation}
is the average over all $|\psi_0\rangle$ of the quadratic entanglement entropy  $E^{(2)}(S,T)$ of the associated history state, with 
the integral running over the whole set of initial states $|\psi_0\rangle$ with the Haar measure $d\psi_0$ \cite{BR.18}. 
We have verified this relation by considering the full set of 52 available polarization unitaries extracted from the experimental characterization, which provided a value $ E^{(2)}({\cal W})=0.712$. A simulation with 1000 random initial states  satisfied the previous relation with and error less than 0.01. 

\section{Conclusions}\label{conclusions}
We have presented a simple optical implementation for realizing discrete  history states. The approach is based on the entanglement between the polarization and spatial DOFs generated by the SLM, and can be used to generate history states with a controllable number of time steps for a qubit system. It  enables  an efficient determination of time averages through a single measurement. The experimental results obtained with the previous scheme show in fact an excellent agreement between both, the direct and sequential method, and also with the theoretical results.
The associated ``system-clock'' entanglement, which is a measure of the distinguishability of the evolved polarization states,  was also determined and shown to characterize the basic features of the discrete trajectories obtained for different initial states. The entangling power of the setup, which determines the average quadratic entanglement that it generates when applied to random initial states, was also analyzed.  
Variations of the present scheme based on two entangled photons could provide a  realization of discrete history states of higher dimensional systems, and are currently under development. 

\vskip 0.5cm
\acknowledgments
\vskip -0.2cm
We express our gratitude to Prof. C. T. Schmiegelow for providing us with the sCMOS camera. This work was supported by the Agencia Nacional de Promoci\'on de Ciencia y T\'ecnica ANPCyT (PICT 2014-2432) and Universidad de Buenos Aires (UBACyT 20020170100040B). R. R. acknowledges support from CIC of Argentina. 

\bibliography{biblio}

\end{document}